\documentstyle[12pt]{article}
\def\beq{\begin{equation}}
\def\eeq{\end{equation}}
\def\bea{\begin{eqnarray}}
\def\eea{\end{eqnarray}}
\def\bq{\begin{quote}}
\def\eq{\end{quote}}

\def\PLB{{\it Phys. Lett.} }

\def\PR{{\it Phys.Rev.} }

\def\gappeq{\mathrel{\rlap {\raise.5ex\hbox{$>$}}
{\lower.5ex\hbox{$\sim$}}}}

\def\lappeq{\mathrel{\rlap{\raise.5ex\hbox{$<$}}
{\lower.5ex\hbox{$\sim$}}}}

\begin{document}
\topmargin -0.5cm
\oddsidemargin -0.3cm
\evensidemargin -0.8cm
\pagestyle{empty}
\begin{flushright}
{revised CERN-TH.7400/94}\\
{hep-ph/9408395}
\end{flushright}
\vspace*{5mm}
\begin{center}
{\bf ESTIMATES OF THE HIGHER-ORDER QCD
 CORRECTIONS:}\\  {\bf THEORY AND APPLICATIONS} \\
\vspace*{1cm}
{\bf Andrei L. Kataev} \footnote{On leave of absence from Institute for
Nuclear Research of the Russian Academy of Sciences, 117312 Moscow,
Russia} \\
\vspace{0.3cm}
Theoretical Physics Division, CERN, \\
CH-1211 Geneva 23, Switzerland \\
\vspace{0.5cm}
and \\
\vspace*{0.5cm}
{\bf Valery V. Starshenko} \\
\vspace*{0.3cm}
Chair of General Physics, \\
Zaporozhye Industrial Institute \\
330600 Zaporozhye, Ukraine \\
\vspace*{1cm}
{\bf ABSTRACT} \\
\end{center}
\vspace*{2mm}
\noindent
We consider the further development of the formalism of the
estimates of  higher-order perturbative corrections in
the Euclidean region, which is based on the application of
the scheme-invariant methods, namely the principle of minimal
sensitivity and the effective charges approach.
We present the estimates of the order $O(\alpha^{4}_{s})$ QCD
corrections to
the Euclidean quantities: the $e^+e^-$-annihilation $D$-function
and the
deep inelastic scattering sum rules, namely  the
non-polarized and polarized Bjorken sum rules and to the Gross--Llewellyn
Smith sum rule.
 The results for the
$D$-function are further applied to estimate the $O(\alpha_s^4)$
QCD corrections to the Minkowskian quantities
$R(s) = \sigma_{tot} (e^{+}e^{-} \to {\rm hadrons}) /
\sigma (e^{+}e^{-} \to \mu^{+} \mu^{-})$ and
$R_{\tau} = \Gamma (\tau \to \nu_{\tau} + {\rm hadrons}) /
\Gamma (\tau \to \nu_{\tau} \overline{\nu}_{e} e)$.
The problem of the fixation of the uncertainties due to the
$O(\alpha_s^5)$ corrections to the considered quantities is also
discussed.

\vspace*{1cm}
\noindent{Shortened vesrion to
submitted to the Proceedings of the QCD-94 Workshop,
Montpellier, France, 7-13 July 1994, to be published in
Nucl. Phys. B Proc. Suppl., ed. S. Narison}

\begin{flushleft} revised CERN-TH.7400/94 \\
August-September 1994.
\end{flushleft}


\newpage
\setcounter{page}{1}
\pagestyle{plain}

{\bf I.}~~~
The main aim of this work, which is based on the considerations
 of Refs. \cite{KatSt1,KatSt2},
 is to obtain the estimates of the higher-order QCD
 corrections to $R(s)$, $R_{\tau}$ and deep-inelastic scattering
sum rules using the improvement formula \cite{rr} of re-expansion
of the expressions for the physical quantities
obtained within the principle of minimal sensitivity (PMS)
\cite{rr} and the
effective charges (ECH) approach \cite{uu}, which is equivalent {\it a
posteriori} to  scheme-invariant perturbation theory \cite{vv}.

Of course, this approach to the estimation of the values of the
uncalculated higher-order terms can not be considered as an alternative
to  direct analytical or numerical calculations. However, we hope
that this method can give an idea of the possible values
of the higher-order terms. This hope is based on the fact that the
application of this method for the estimate of the four-loop corrections
to the expression for $(g-2)_e$ \cite{ss} gave  results which
turned out to be in  surprisingly good agreement with the latest
results of the direct numerical calculations of Ref. \cite{Kin}
\footnote{A detailed re-consideration of the analysis of Ref.
\cite{ss} and its generalization to  five-loop order will be
presented in a more detailed work \cite{fut}.}. The second
argument in favour of this procedure came from its successful
application for the analysis of the Drell-Yan
cross section at the $O(\alpha_s^2)$-level \cite{Aurenche}.
 Moreover, as
will be demonstrated in our work, the re-expansion formalism of
Ref. \cite{rr} is also working quite well in QCD at the $NNLO$
for at least three independent quantities, namely for the
$e^+e^-$-annihilation $D$-function and the non-polarized and polarized
Bjorken sum rules. Therefore, we prefer to consider these
\underline{five} facts not as a surprising accidental coincidence,
but as an argument in favour of more detailed studies of the
intrinsic features of this approach and its various applications.

In this work we further develop the re-expansion formalism of Ref.
\cite{rr}, deriving  new terms in the corresponding improvement
formula. The previously-known terms are used to obtain the estimates
of the next-to-next-to-next-to-leading order ($N^3LO$) QCD
 corrections to the $D$-function, the non-polarized
Bjorken sum rule and polarized Bjorken sum rule, which is closely
related by the structure of the corresponding perturbative series
to the Gross--Llewellyn Smith sum rule.
 We will  use the
results obtained for the $D$-function to estimate the effects of the
$N^3LO$ corrections to the perturbative series for the Minkowskian
quantities $R(s)$ and $R_{\tau}$ by adding the explicitly
calculable terms, previously discussed in
Refs.\cite{Bj,pp}, supplementing thus the related
considerations of Refs. \cite{rr}, \cite{ww}
by the additional input information.  The derived
new term in the improvement formula is applied
to touch on the problem of fixing
the values of the $O(\alpha_s^5)$-corrections to the analysed
quantities.

{\bf II.}~~~
Consider first the order $O(a^{N})$ approximation of a Euclidean
physical quantity
\beq
D_{N} = d_{0} a(1 + \sum^{N-1}_{i=1} \, d_{i} a^{i})
\label{1}
\eeq
with $a = \alpha_{s}/\pi$ being the solution of the corresponding
renormalization group equation for the $\beta$-function which is
defined as
\beq
\mu^{2} \frac{\partial a}{\partial \mu^{2}} =
\beta (a) = - \beta_{0} a^{2}
(1 +  \sum^{N-1}_{i=1} \, c_{i} a^{i})\ .
\label{2}
\eeq
In the process of the concrete calculations of the coefficients
$d_{i}, i \geq 1$ and $c_{i}, i \geq 2$, the $\overline{MS}$ scheme is
commonly used.  However, this scheme is not the unique prescription for
fixing the RS ambiguities (for the recent discussions see e.g., Ref.
\cite{ww}).

The PMS \cite{rr} and ECH \cite{uu} prescriptions stand out from
various methods of treating scheme-dependence ambiguities.
 Indeed, they are based on the conceptions of the
scheme-invariant quantities, which are defined as the combinations of
the scheme-dependent coefficients in Eqs. (1) and (2).  Both these
methods pretend to be
 the ``optimal" prescriptions, in the sense that they might
 provide better convergence of the corresponding approximations in
the non-asymptotic regime, and thus allow an estimation of the
uncertainties of the perturbative series
in the definite order of perturbation theory.  Therefore,
applying these
 ``optimal" methods,  one can try to estimate the
effects of the order $O(a^{N+1})$-corrections starting from the
approximations $D^{opt}_{N} (a_{opt})$ calculated in a certain
``optimal" approach \cite{rr}, \cite{ss}, \cite{tt}.

Let us following the considerations of Ref. \cite{rr} and
re-expand $D_{N}^{opt} (a_{opt})$ in terms of the coupling
constant $a$ of the particular scheme
\beq
D_{N}^{opt} (a_{opt}) = D_{N} (a) + \delta D_{N}^{opt} a^{N+1}
\label{3}
\eeq
where
\beq
\delta D_{N}^{opt} = \Omega_{N}(d_{i}, c_{i}) +
\Omega_{N} (d_{i}^{opt}, c_{i}^{opt})
\label{4}
\eeq
are the  numbers which simulate the coefficients of the
order $O(a^{N+1})$-corrections to the physical quantity, calculated in
the particular initial scheme, say the $\overline{MS}$-scheme.

The explicit form of the coefficients $\Omega_{i}$
can be obtained
following the considerations of Ref. \cite{rr}.  We
present here  the final already known expressions \cite{rr}:
\beq
\Omega_{2} = d_{0}d_{1} (c_{1} + d_{1}),
\label{6}
\eeq
\beq
\Omega_{3} = d_{0}d_{1} (c_{2} - \frac{1}{2} c_{1}d_{1}
-2d_{1}^{2} + 3d_{2})\ .
\label{7}
\eeq
and the new  term $\Omega_{4}$ evaluated by us:
\bea
\Omega_{4} =\frac{d_0}{3} ( 3c_{3}d_1+c_2d_2-4c_2d_1^2+2c_1d_1d_2
-c_1d_3 +14d_1^4-28d_1^2d_2+5d_2^2+12d_1d_3 )
\label{new1}
\eea
which reproduces the renormalization-group controllable logarithmic
terms at the five-loop level \cite{Kat1}.

It should be stressed that in the ECH approach $d_{i}^{ECH} \equiv 0$
for all $i \geq 2$.  Therefore one gets the following expressions for
the $NNLO$ and $N^3LO$ corrections in Eq. (3):
\beq
\delta D_{2}^{ECH} = \Omega_{2} (d_{1}, c_{1})
\label{8}
\eeq
\beq
\delta D_{3}^{ECH} = \Omega_{3}(d_{1}, d_{2}, c_{1}, c_{2})
\label{9}
\eeq
\beq
\delta D_{4}^{ECH}=\Omega_4(d_1,d_2,d_3,c_1,c_2,c_3)
\label{new2}
\eeq

In order to find similar corrections to Eq.(3) in the N-th order of
perturbation theory starting from the PMS approach \cite{rr}, it is
necessary to use the relations obtained in Ref. \cite{yy} between the
coefficients $r_{i}^{PMS}$ and $c_{i}^{PMS} \; (i \geq 1)$ in the
expression for the order $O(a^{N}_{PMS})$ approximation
$D^{PMS}_{N} (a_{PMS})$ of the physical quantity under consideration.
The corresponding corrections have the following form:
\beq
\delta D_{2}^{PMS} = \delta D_{2}^{ECH} + \frac{d_{0} c^{2}_{1}}{4}
\label{10}
\eeq
\beq
\delta D_{3}^{PMS} = \delta D_{3}^{ECH}\ .
\label{11}
\eeq
Notice the identical coincidence of the $N^3LO$ corrections obtained
starting from both the PMS and ECH approaches.  A similar observation
was made in Ref. \cite{ss} using different (but related) considerations.

In the fourth order of  perturbation theory the additional
contribution to $\delta D_{4}^{PMS}$ has more complicated structure.
Indeed, the expression for $\Omega_4(d_i^{PMS},c_i^{PMS})$ in
Eq. (4) reads:
\bea
\Omega_4(d_i^{PMS},c_i^{PMS})=\frac{d_0}{3} [ \frac{1}{4}c_1c_3^{PMS}
 -\frac{4}{81}(c_2^{PMS})^2
-\frac{5}{81}c_1^2c_2^{PMS}+\frac{7}{648}c_1^4]
\label{omega4}
\eea
where
\beq
c_2^{PMS}=\frac{9}{8}(d_2+c_2-d_1^2-c_1d_1+\frac{7}{36}c_1^2)
+O(a)
\label{c2pms}
\eeq
and
\bea
c_3^{PMS}=4(d_3+\frac{1}{2}c_3-c_2d_1-3d_1d_2+2d_1^3)
+\frac{1}{2}c_1(d_2+c_2+3d_1^2-c_1d_1+\frac{1}{108}c_1^2)+O(a)
\label{c3pms}.
\eea
The expressions for Eqs. (14)- (16) are  pure numbers, which
do not depend on the choice of the initial scheme.
Note, that we have checked that in the case of the consideration
of the perturbative series for $(g-2)_e$ and $(g-2)_{\mu}$
the numerical values of $\Omega_4(d_i^{PMS},c_i^{PMS})$ are
small and thus the {\it a posteriori} approximate equivalence
of the ECH and PMS approaches is preserved for the quantities
under consideration at this level also \cite{fut}. We think that
this feature is also true in QCD.

{\bf III.}~~~
Consider now   the familiar characteristic of the $e^{+}e^{-} \to
\gamma \to {\rm hadrons}$ process, namely the $D$-function defined in
the Euclidean region:
\beq
D(Q^{2}) = Q^{2} \int^{\infty}_{0} \, \frac{R(s)}{(s+Q^{2})^{2}} \, ds
\label{12}
\eeq
Its perturbative expansion has the following form:
\begin{eqnarray}
D(Q^{2}) &=& 3 \Sigma Q^{2}_{f} [1 + a + \sum _{i\geq 1} \,
 d_{i}a^{i+1}  ]
+  (\Sigma Q_{f})^{2} [ \tilde{d}_{2} a^{3} + O(a^{4}) ]
\label{13}
\eea
where $Q_{f}$ are the quark charges, and the structure proportional to
$(\Sigma Q_{f})^{2}$ comes from the light-by-light-type diagrams. The
coefficients $d_{1}$ and $d_{2}, \tilde{d}_{2}$ were calculated in the
$\overline{MS}$-scheme in Refs. \cite{ggg} and \cite{aaa,bb}
 respectively.
%
Following the proposals of Ref. \cite{aai}, we will treat the
light-by-light-type term in Eq. (18) separately from the ``main"
structure of the $D$-function, which is proportional to the
quark-parton expression $D^{QP}(Q^{2}) = 3 \Sigma Q^{2}_{f}$. In fact,
one can hardly expect that it is possible to predict higher-order
coefficients $\tilde{d}_{i}, i \geq 3$ of the second structure in Eq.
(17) using the only explicitly-known term $\tilde{d}_{2}$.  Therefore
we will neglect the light-by-light-type structure as a whole in all our
further considerations.  This approximation is supported by the
relatively tiny contribution of the second structure of Eq. (17) to the
final $NNLO$ correction to the $D$-function.

The next important ingredient of our analysis is the QCD
$\beta$-function (2), which is known in the MS-like schemes at the $NNLO$
level \cite{bbi}.

Using now the perturbative expression for the $D$-function, one can
obtain the perturbative expression for $R(s)$, namely
\begin{equation}
R(s)  =  3 \Sigma Q^{2}_{f} [1 + a_{s} + \sum_{i\geq 1} \,
r_{i} a^{i+1}_{s} ]
 +  (\Sigma Q_{f})^{2} [ \tilde{r}_{2} a^{3}_{s} + ... ]
\label{17}
\end{equation}
where $a_{s} = \bar{\alpha}_{s} / \pi$,
$r_{1}  =  d_{1}$,
$r_{2} =  d_{2} - \pi^{2} \beta^{2}_{0}/3$,
$\tilde{r}_{2} = \tilde{d}_{2}$ and
$r_{3}  =  d_{3} - \pi^{2} \beta^{2}_{0}
(d_{1} + \frac{5}{6} c_{1})$ .
The corresponding $\pi^{2}$-terms come from the analytic continuation
of the Euclidean result for the $D$-function to the physical region.
The effects of the higher-order $\pi^{2}$-terms were discussed in
detail in Ref. \cite{Bj}. For example, the corresponding expression for
the $r_4$-term reads :
\bea
r_{4} = d_{4} -\pi^2\beta_0^2 (2d_2+\frac{7}{3}c_1d_1
+\frac{1}{2}c_1^2+c_2)+\frac{\pi^4}{5}\beta_0^4
\label{r4}
\eea

The perturbative expression for $R_{\tau}$ is defined as
\bea
R_{\tau}  =  2 \int_{0}^{M^{2}_{\tau}} \, \frac{ds}{M^{2}_{\tau}} \,
(1 - s/M^{2}_{\tau})^{2} \, (1 + 2s/M^{2}_{\tau}) \tilde{R}(s)
 \simeq  3[1 + a_{\tau} + \sum_{i\geq 1} \,
r_{i}^{\tau} a^{i+1}_{\tau} ]
\label{19}
\eea
where $a_{\tau} = \alpha_{s} (M^{2}_{\tau}) / \pi$ and $\tilde{R} (s)$
is $R(s)$ with
with $f = 3, (\Sigma Q_{f})^{2} = 0, 3 \Sigma Q^{2}_{f}$ substituted
for $3 \Sigma \mid V_{ff'} \mid^{2}$ and $\mid V_{ud} \mid^{2} +
\mid V_{us} \mid^{2} \approx 1$.

It was shown in Ref. \cite{pp} that it is convenient to
express the coefficients of the
series (24) through  those ones of the series (18) for the
$D$-function in the following form :
\bea
r^{\tau}_{1} & = &
d_{1}^{\overline{MS}} (f = 3) +  g_1   \nonumber \\
r^{\tau}_{2} & = &
d_{2}^{\overline{MS}} (f = 3) +  g_2  \nonumber
\\
r^{\tau}_{3} & = &
d_{3}^{\overline{MS}} (f = 3) + g_3
\label{20}
\eea
where in our notations
\bea
g_1 & = & -\beta_0I_1 = 3.563 \nonumber \\
g_2 & = & -[2d_1+c_1]\beta_0 I_1+\beta_0^2 I_2=19.99 \nonumber \\
g_3 & = & -[3d_2+2d_1c_1+c_2]\beta_0 I_1+
[3d_1+\frac{5}{2}c_1]\beta_0^2 I_2
-\beta_0^3 I_3=78.00
\label{gn}
\eea
and $I_{k}$ are defined and calculated
in Ref. \cite{pp}. Their  analytical expressions
read: $I_1$=-19/12, $I_2$=265/72-$\pi^2/3$ and $I_3$=-3355/288+
19$\pi^2$/12. One of the pleasant features of Eqs.(\ref{gn}) is that
they are absorbing all effects of the analytical continuation.

Following the lines of Ref.\cite{pp} we derive the corresponding
expression for the coefficient $r^{\tau}_{4}$:
\bea
r^{\tau}_{4} = d_4^{\overline{MS}}(f=3) + g_4
\label{r4tau}
\eea
where
\begin{eqnarray}
g_4&=&-[4d_3+3d_2c_1+2d_1c_2+c_3]\beta_0I_1+[6d_2+7c_1d_1+
\frac{3}{2}c_1^2+3c_2]\beta_0^2I_2   \nonumber \\
&& -[4d_1+\frac{13}{3}c_1]\beta_0^3I_3
+\beta_0^4I_4 = 3.562c_3(f=3)+14.247d_3(f=3)-466.818
\label{g4}
\end{eqnarray}
and $I_4$=41041/864-265$\pi^2$/36+$\pi^4$/5 $\approx -5.668$.

In order to estimate the values of the order $O(a^{3})$, $O(a^{4})$
and $O(a^5)$
corrections to $R(s)$ and $R_{\tau}$, we will apply Eqs. (5) - (10) in
the Euclidean region to the perturbative series for the $D$-function
and then obtain the  estimates we are interested in
using Eqs. (18), (19), (21)-(24).
This is the new ingredient of this analysis, which distinguishes
it from the related ones of Refs. \cite{rr}, \cite{ww}.

We now recall the perturbative expression for the non-polarized Bjorken
deep-inelastic scattering sum rule
\bea
BjnSR  = \int^{1}_{0} \, F_{1}^{\bar{v}p - vp} (x, Q^{2}) dx
 =  1 - \frac{2}{3} a (1 + \sum_{i\geq 1} \,
 d_{i}a^i )
\label{21}
\eea
where the coefficients $d_{1}$ and $d_{2}$ are known in the
$\overline{MS}$
scheme from the results of calculations of Ref. \cite{jj} and Ref.
\cite{ee} .
The expression for the polarized Bjorken sum rule BjpSR has the
following form:
\bea
BjpSR  = \int^{1}_{0} \, g_{1}^{ep-en} (x, Q^{2}) dx
 =  \frac{1}{3} \mid \frac{g_A}{g_V} \mid \,
[ 1-a (1 + \sum_{i\geq 1} \, d_{i} a^{i}) ]
\label{24}
\eea
where the coefficients $d_{1}$ and $d_{2}$ were explicitly calculated
in the $\overline{MS}$ scheme in
Refs. \cite{kk} and \cite{ff} respectively.

It is also worth emphasizing that, in spite of the identical
coincidence of the $NLO$ correction to the Gross-Llewellyn Smith sum rule
$GLSSR  =  (1/2)\int^{1}_{0} F_3^{\overline{\nu}p+\nu p}
(x,Q^2)dx$
with the ones for the $BjpSR$ \cite{kk}
 the corresponding $NNLO$
 correction
differs from the one     for the $BjpSR$
 by the contributions of the
light-by-light-type terms typical of the GLSSR \cite{ff}.
Since these light-by-light-type
terms appear for the first time at the $NNLO$, it is impossible to
predict the value of the light-by-light-type contribution at the $N^3LO$
level using the corresponding $NNLO$ terms as the input information.
However, noticing that at the $NNLO$ level
the corresponding light-by-light-type contributions are small \cite{ff},
we will assume that the similar contributions
are small at the $N^3LO$ level also. Only after this assumption can
our estimates of the $NNLO$ and $N^3LO$ corrections to the BjpSR  be
considered also as the estimates of the corresponding
corrections in the perturbative series for the GLSSR .

{\bf IV.}~
The estimates of the coefficients of the order $O(a^{3})$,
$O(a^{4})$ and $O(a^{5})$
QCD corrections to the $D$-function, $R(s)$, BjnSR and
BjpSR/GLSSR obtained from the ECH-improved expressions
 are presented in Tables 1 - 4
respectively.  Due to the complicated $f$-dependence of the
coefficients $\Omega_{2}, \Omega_{3}$ , $\Omega_{4}$
in Eqs. (5)-(7), we are
unable to predict the explicit $f$-dependence of the corresponding
coefficients in the form respected by perturbation theory.  The
results are presented for the fixed number of quark flavours $1 \leq f
\leq 6$ and are normalized to the $\overline{MS}$-scheme.
\begin{center}
\begin{tabular}{|c|c|c|c|c|} \hline
$f$ & $d^{ex}_{2}$ & $d^{est}_{2}$ &
$d^{est}_{3}$ & $ d_4^{est}-c_3d_1$\\ \hline
1 & 14.11 & 7.54 &  75 & 474 \\ \hline
2 & 10.16 & 6.57 &  50 & 261 \\  \hline
3 & 6.37 & 5.61 &  27.5 & 111 \\ \hline
4 & 2.76 & 4.68 &  8  & 23 \\  \hline
5 & -0.69 & 3.77 &  -8 & -15  \\ \hline
6 & -3.96 & 2.88 &  -21 & -1.8\\ \hline
\end{tabular}
\end{center}
\vspace*{0.5mm}
Table 1 : The results of estimates of the $NNLO$, $N^3LO$ and $N^4LO$
 corrections in
the series for the
$D$-functions.
\begin{center}
\begin{tabular}{|c|c|c|c|c|} \hline
$f$ & $r^{ex}_{2}$ & $r^{est}_{2}$ &
$r^{est}_{3}$ & $r_4^{est}-c_3d_1$\\ \hline
1 & -7.84 & -14.41 &  -166 & -1748\\ \hline
2 & -9.04 & -12.63 &  -147 & -1156 \\ \hline
3 & -10.27 & -11.03 &  -128 & -669 \\ \hline
4 & -11.52 & -9.58 &  -112 & -263 \\ \hline
5 & -12.76 & -8.29 &  -97 & 64 \\ \hline
6 & -14.01 & -7.17 &  -83 & 334 \\ \hline
\end{tabular}
\end{center}
\vspace*{0.5mm}
Table 2: The results of estimates of the $NNLO$, $N^3LO$ and $N^4LO$
 corrections in
the series for $R(s)$.
\newpage
\begin{center}
\begin{tabular}{|c|c|c|c|c|} \hline
$f$ & $d^{ex}_{2}$ & $d^{est}_{2}$ &
$d^{est}_{3}$ & $d_4^{est}-c_3d_1$ \\ \hline
1 & 44.97 & 39.62 &  424 & 4177 \\ \hline
2 & 36.19 & 33.28 &  303 & 2619 \\ \hline
3 & 27.89 & 27.37 &  200 & 1481 \\ \hline
4 & 20.07 & 21.91 &  114 & 671 \\ \hline
5 & 12.72 & 16.91 &  44 & 137 \\ \hline
6 & 5.85 & 12.39 &  -10 & -148 \\ \hline
\end{tabular}
\end{center}
\vspace*{0.5mm}
Table 3: The results of estimates of the $NNLO$, $N^3LO$ and $N^4LO$
 corrections in
the series for BjnSR.
\begin{center}
\begin{tabular}{|c|c|c|c|c|} \hline
$f$ & $d^{ex}_{2}$ & $d^{est}_{2}$ &
$d^{est}_{3}$ & $d_4^{est}-c_3d_1$\\ \hline
1 & 34.01 & 27.25 &  290 & 2557\\ \hline
2 & 26.93 & 23.11 &  203 & 1572 \\ \hline
3 & 20.21 & 19.22 &  130 & 854\\ \hline
4 & 13.84 & 15.57 &  68 & 342 \\ \hline
5 & 7.83 & 12.19 &  18 & 27 \\ \hline
6 & 2.17 & 9.08 &  -22 & -135 \\ \hline
\end{tabular}
\end{center}
\vspace*{0.5mm}
Table 4: The results of estimates of the $NNLO$, $N^3LO$ and $N^4LO$
 corrections in the series for BjpSR and GLSSR.

 The presented $NNLO$ estimates
are in reasonable  agreement with the results of
 the explicit calculations. The best agreement is achieved for
$f=3$ numbers of flavours. It can be shown that
the difference with the results obtained
staring from the PMS improved expression is rather small
\cite{KatSt1,KatSt2}.

In order to address the problem of fixing the values of the
$O(a^5)$ QCD corrections to the considered Euclidean quantities,
 Eqs. (7), (10) with  explicitly calculated $NLO$
and $NNLO$ coefficients $d_1,d_2,c_1,c_2$ were applied. The
values of the coefficients $d_3$ were fixed by using the
estimates of the $N^3LO$ coefficients
 the determined
from Eqs. (6), (9).
To  estimate the next-to-next-to-next-to-next-to-leading order
($N^4LO$) coefficient of $R(s)$,
which is  related to the $N^4LO$ coefficient $d_4$ of the $D$-function,
the explicitly calculated terms in Eq.(19) were taken
into account.
However, the expressions for $\Omega_4$ in Eq. (7)
 depend also on the  four-loop
coefficient $c_3$ of the QCD $\beta$-function which is unknown
at present. Therefore, in
Tables 1-4 we presented the estimates
 of the combinations
$d_4-d_1c_3$ and $r_4-r_1c_3$. Another  existing uncertainty in the
value of $N^3LO$ coefficients $d_3$ is fixed by the assumption
that the real values of these coefficients do not significantly
differ from the  $N^3LO$ estimates obtained by us.
 However, in
the case  of the
$D$-function with $f=3$ numbers of flavours we present the
more detailed expression of $d_4$:
\beq
d_4^{est}(f=3)= 1.64 c_3 (f=3) + 5.97 d_3(f=3)-52.66  .
\label{de}
\eeq
In order to obtain the corresponding estimate from Table 1, Eq.(27)
should be supplemented by the $N^3LO$-estimate $d_3^{est}(f=3)=27.5$
first obtained in Ref.\cite{KatSt1}. This result is in good agreement
with the ``geometric progression'' assumption $d_3(f=3)
\approx d_2^2(f=3)/d_1(f=3) \approx 25$ of Ref.\cite{pp}. Our result
was recently supported \cite{Dib} by the phenomenological analysis
of the ALEPH data for $R_{\tau}$.

Amongst other  most interesting results of our analysis
are the estimates of the $N^3LO$ and $N^4LO$
corrections to
 $R(s)$  for $f = 5$ numbers of flavours.  Taking $\alpha_{s}
(M_{Z}) \approx 0.12$, we get the estimate of the corresponding $N^3LO$
contribution to $\Gamma (Z^{0} \to {\rm hadrons})$
\beq
(\delta \Gamma_{Z^{0}})_{N^3LO}
\approx -97 (a(M_{Z}))^{4} \approx -2 \times
10^{-4}\ .
\label{32}
\eeq
The comparison of the results of Table 1 with those of Table 2
demonstrate that the $\pi^2$-effects give dominating contributions
to the coefficients of $R(s)$ and $\Gamma_{Z^{0}}$. Indeed, at the
$N^4LO$-level they are also larger than the estimated coefficient
$d_4$ of the Euclidean $D$-function. Assuming that
$c_{3}(f=5)=c_{2}(f=5)^2/c_{1}(f=5)\approx 1.715$ we arrive at
the following less substantiated  estimate of
 the corresponding $N^4LO$ contributions, namely
\beq
(\delta \Gamma_{Z^{0}})_{N^4LO}
\approx 64 (a(M_{Z}))^{5} \approx 5 \times
10^{-6}
\label{znew}
\eeq
The  way of fixation of the value
of $c_3$ used above
is not applicable for the case of $f=6$, since we expect
that in this case its real value is negative.

Using now the relations
derived in Refs. \cite{BPN,pp}  between
the coefficients of the $D$-function and $R_{\tau}$ and taking
$\alpha_{s}(M_{\tau})\approx 0.36$ \cite{BPN} we get the
following estimate
 of the $N^3LO$ contribution to $R_{\tau}$:
\beq
(\delta R_{\tau})_{N^3LO} \approx 105.5 \, a^{4}_{\tau}
\approx 1.8 \times 10^{-2}\ .
\label{33}
\eeq
This estimate is more precise than
those presented in Refs. \cite{BPN} and \cite{pp}, namely $\delta
R_{\tau} = \pm 130 \, a^{4}_{\tau}$ \cite{BPN} and $\delta R_{\tau} =
(78 \pm 25) a^{4}_{\tau}$ \cite{pp},
and is smaller than
the result of applying the Pad\'e resummation technique directly to
$R_{\tau}$, namely $\delta R_{\tau} = 133 a^{4}_{\tau}$ \cite{sli}.

Generalizing the considerations of Ref.\cite{pp} to the
order $O(a_{\tau}^5)$-level  and taking
$c_3(f=3)\approx c_2(f=3)^2/c_1(f=3) \approx 11$ we get the
estimate   of the $N^4LO$ corrections to $R_{\tau}$ \cite{KatSt2}. This
estimate reads \cite{KatSt2}:
\beq
(\delta R_{\tau})_{N^4LO} \approx 94  a^{5}_{\tau}
\approx 1.8 \times 10^{-3}\ .
\label{33}
\eeq

{\bf V.}~~
Let us emphasize that
 the results for $R(s)$ and
$R_{\tau}$ were obtained by us
from those    of the $D$-function after
taking into account explicitly calculable terms of the analytic
continuation to the Minkowskian region, discussed in detail in
Ref. \cite{Piv}. In principle, one can
try to study the application of the procedure used by us directly to
$R(s)$ and $R_{\tau}$ in the Minkowskian region. We checked that
in the case of applications of Eqs. (5)-(7) with the substitution
of $d_i=r_i$ or $d_i=r_i^{\tau}$
one can reproduce the same values of the Eucledian
coefficients $d_i$ but the structure of the high order $\pi^2$-terms
will not agree with the explicitly known results.
A similar problem will also definitely arise in the case
of more rigorous studies of the applicability of the Pad\'e
resummation methods for the Minkowskian quantities $R(s)$ and
$R_{\tau}$ (for the  studies existing at present see Refs.
\cite{sli}). We think that these
analyses should be supplemented by the construction of the
Pad\'e approximants directly to the $D$-function.
In our case it is possible to reproduce our results
$R(s)$ and $R_{\tau}$ by means of application of
 a similar technique directly in the Minkowskian region
after modifications of
Eqs. (5)-(7) in the Minkowskian region by adding concrete calculable
$\pi^2$-dependent and scheme-independent factors.

{\bf Acknowledgements}

We are grateful to R.N. Faustov for attracting our attention to the
necessity for    detailed consideration of
 the results of Ref. \cite{ss} at the
preliminary stage  of our similar QED studies, which will be presented
elsewhere, and to J. Ch\'yla, F. Le Diberder and
 G. Marchesini for discussions and to J. Ellis
for the useful comments on the earlier version of this paper.

\end{document}